\documentclass[twocolumn,showpacs]{revtex4}

\usepackage{graphicx}
\usepackage{dcolumn}
\usepackage{bm}
\usepackage{hyperref}

\begin{document}

\title{A Simple method to set up low eccentricity initial data 
       for moving puncture simulations}

\author{Wolfgang Tichy}
\affiliation{Department of Physics,
	Florida Atlantic University, 
	Boca Raton, FL 33431, USA}
\author{Pedro Marronetti}
\affiliation{Department of Physics,
	Florida Atlantic University, 
	Boca Raton, FL 33431, USA}


\pacs{
04.25.dg,	
04.25.Nx,	
04.20.Ex,	
04.30.Db,	
}


%
\newcommand\be{\begin{equation}}
\newcommand\ba{\begin{eqnarray}}

\newcommand\ee{\end{equation}}
\newcommand\ea{\end{eqnarray}}
\newcommand\p{{\partial}}
\newcommand\remove{{{\bf{THIS FIG. OR EQS. COULD BE REMOVED}}}}
%

\begin{abstract}

We introduce two new eccentricity measures to analyze numerical simulations.
Unlike earlier definitions these eccentricity measures do not involve any
free parameters which makes them easy to use.
We show how relatively inexpensive grid setups can be used to estimate
the eccentricity during the early inspiral phase.
Furthermore, we compare standard puncture data and post-Newtonian
data in ADMTT gauge. We find that both use different coordinates.
Thus low eccentricity initial momentum parameters for a
certain separation measured in ADMTT coordinates are hard to use
in puncture data, because it is not known how the separation in puncture
coordinates is related to the separation in ADMTT coordinates.
As a remedy we provide a simple approach which allows us to iterate 
the momentum parameters until our numerical simulations result
in acceptably low eccentricities.

\end{abstract}

\maketitle

\section{Introduction}

Currently several gravitational wave detectors such as 
LIGO~\cite{LIGO:2007kva,LIGO_web},
Virgo~\cite{VIRGO_FAcernese_etal2008,VIRGO_web}
or GEO~\cite{GEO_web} are already operating,
while several others are in the
planning or construction phase~\cite{Schutz99}. One of the most promising
sources for these detectors are the inspirals and mergers of binary black
holes. In order to make predictions about the final phase of such
inspirals and mergers, fully non-linear numerical simulations of the Einstein
Equations are required.
To numerically evolve the Einstein equations, at least
two ingredients are necessary. First we need a specific formulation
of the evolution equations. And second, to start such simulations initial
data are needed. As the first ingredient most groups nowadays
use the BSSNOK formulation~\cite{Nakamura87,Shibata95,Baumgarte:1998te}
of the evolution equations.
This formulation is usually evolved using finite differencing methods,
but for single black holes there have been some attempts to use
spectral methods~\cite{Tichy:2006qn,Tichy:2009yr,Tichy:2009zr}.
For binary black holes the BSSNOK system is usually used together
with the moving puncture approach~\cite{Campanelli:2005dd, Baker:2005vv}.
This approach so far works only with finite differencing methods
since certain evolved variables are not smooth inside the black holes (at the
punctures).
Almost all simulations using the BSSNOK formulation to date use standard
puncture data~\cite{Brandt97b,Ansorg:2004ds} as initial data.
These initial data are very flexible in that they contain free parameters
for the position, momentum and spin of each black hole and thus one can
setup practically any kind of orbit. Note, however,
that the emission of gravitational waves
tends to circularize the orbits~\cite{Peters:1963ux,Peters:1964}.
Thus for realistic binary black hole systems that have been inspiraling already
for a long time, we expect the two black holes to be
in quasi-circular orbits around each other with a radius
which shrinks on a timescale much larger than the orbital
timescale. This means that the initial data should be such that the orbit
has no or at least very small eccentricity. For our purposes
here we follow the 
NRAR (Numerical Relativity - Analytical Relativity)
collaboration~\cite{NRAR_web} guidelines
which consider eccentricities of order a few times $10^{-3}$
acceptably small.
There have been several previous works that have considered 
eccentricities for puncture initial 
data~\cite{Baker:2006ha,Husa:2007rh,Walther:2009ng}. However, the most 
successful approach in terms of achieving low eccentricities
was implemented for excision type initial 
data~\cite{Pfeiffer:2007yz,Boyle:2007ft,Mroue:2010re}.
The method discussed in this work aims at lowering the
eccentricity for the kind of puncture initial data that is routinely used
with the moving puncture approach.

Throughout we will use units where $G=c=1$. The black hole masses
are denoted by $m_1$ and $m_2$. We also introduce the total mass
$M=m_1+m_2$, the reduced mass $\mu = m_1 m_2/M$ and $\nu=\mu/M$.

The paper is organized as follows. 
Sec.~\ref{eccentricity} introduces and compares several eccentricity
measures. 
In Sec.~\ref{numerics} we describe grid setups that can be used in
numerical simulations aimed at measuring the eccentricity.
Sec.~\ref{parchoices} discusses a simple method to pick initial
momentum parameters. This is followed by Sec.~\ref{ecc_reduc}
which describes how to iterate these parameters to arrive at
a reduced eccentricity.
We conclude with a discussion of our results in Sec.~\ref{discussion}.

\section{Defining eccentricity for inspiral orbits}
\label{eccentricity}

Real binary black hole orbits can never be circular. They always follow
spirals. So when we are aiming for low eccentricity initial data, we really
want data that result in trajectories that spiral in smoothly without
oscillations in the black hole separation.
Of course this issue is further complicated by the fact that
trajectories are coordinate dependent.

There are several earlier eccentricity definitions for inspiral orbits
in the literature~\cite{Baker:2006ha,Husa:2007rh,
Pfeiffer:2007yz,Boyle:2007ft,Mroue:2010re,Walther:2009ng}.
All of them define eccentricity as a deviation from an underlying 
smooth, secular trend in some specific quantity that is associated with the
orbits. In~\cite{Baker:2006ha} the frequency of the dominant $l=m=2$ mode
of the gravitational waves emitted
is fitted to a fourth order monotonic polynomial, and the deviation of the
frequency from this fit is used to compute the eccentricity. 
This approach works for non-spinning binaries. Essentially the same method
is also used in~\cite{Husa:2007rh}, but instead of the gravitational wave
frequency~\cite{Husa:2007rh} uses the orbital frequency and also the
coordinate separation to obtain two eccentricity measures. These same
measures are also used in~\cite{Walther:2009ng}.
The approaches in~\cite{Pfeiffer:2007yz,Boyle:2007ft} fit a linear function
plus a sine function to the coordinate separation and also
the proper separation. The eccentricity can then be obtained from the
amplitude of the fitted sine function.
In~\cite{Mroue:2010re} the same fitting approach as in~\cite{Husa:2007rh} is
used, but the fitted quantities are coordinate separation,
proper separation and also orbital frequency.
All the approaches based in orbital parameters should in principle
also work for systems with spin. Also note that all these eccentricity
definitions are chosen such that they result in the correct value
for Newtonian orbits.

Below we will introduce two new eccentricity definitions and compare them
to the earlier definition based on fitting the orbital 
frequency~\cite{Husa:2007rh,Mroue:2010re}.

The first eccentricity definition is based on the coordinate separation
of the two black holes. It is given by
\begin{equation}
\label{e_r_def}
e_{r}(t) = \frac{ \Delta r_{max}(t) - \Delta r_{min}(t) }{2r_{av}}
\end{equation}
where the average separation, and the maximum and minimum deviation from 
a smoothed value $r_{s}$ are given by
\begin{eqnarray}
r_{av} &=& \int_{t-T/2}^{t+T/2} r(t') dt' / T \\
\Delta r_{max}(t) &=& \max_{t'\in[t-T/2,t+T/2]} [r(t')-r_{s}(t',t)] \\
\Delta r_{min}(t) &=& \min_{t'\in[t-T/2,t+T/2]} [r(t')-r_{s}(t',t)] .
\end{eqnarray}
Here the period $T$ is defined using Kepler's law
\begin{equation}
T = 2\pi (r^3/M)^{1/2} .
\end{equation}
Notice that the actual orbital period may be slightly different, but this
estimate suffices to get an approximate eccentricity measure.
The smoothed value $r_{s}(t',t)$ is obtained from
\begin{equation}
\label{r_lin_fit}
r_{s}(t',t) = r(t) + \frac{r(t+T/2)-r(t-T/2)}{T} (t'-t) ,
\end{equation}
but different smoothings are possible (e.g., by performing a 
least-squares fit of e.g. a linear or quadratic function to $r(t)$ in the
interval $[t+T/2,t-T/2]$).
Essentially the definition in Eq.~(\ref{e_r_def}) 
measures how much the coordinate separation oscillates over the time $T$.
For Newtonian orbits it coincides with the usual eccentricity definition
for elliptic orbits. For orbits whose radius shrinks linearly in time
(without any oscillations) $e_{r}(t)$ is zero.

Another similar eccentricity measure can be obtained using the gravitational
wave signal of the inspiraling binary.
The idea is to determine the separation in a more
gauge invariant way from the amplitude of $\Psi_4$ instead of using
the gauge dependent coordinate separation. In~\cite{Buonanno:2006ui}
it is shown that for a non-precessing binary in the quadrupole
approximation the amplitude of the $l=m=2$ spin-weighted Spherical
Harmonic mode is given by
\begin{equation}
|C_{22}| = 32 \sqrt{\pi/5} \nu (M \omega)^{8/3} , 
\end{equation}
where $\omega$ is the orbital angular velocity. Using Kepler's law 
we can define a separation
\begin{equation}
r_{22} = M^{1/3} \omega^{-2/3} 
= M [ |C_{22}|/(32 \sqrt{\pi/5} \nu) ]^{-1/4}
\end{equation}
which is directly related to the amplitude of $|C_{22}|$ of the $l=m=2$ mode
of $\Psi_4$. 
Replacing the coordinate separation in Eq.~(\ref{e_r_def}) by
$r_{22}$ we define
\begin{equation}
\label{e_22_def}
e_{22}(t) = \frac{ \Delta r_{22,max}(t) - \Delta r_{22,min}(t)}{2r_{22,av}} ,
\end{equation}
which is an eccentricity definition that can be computed from $\Psi_4$
alone. Note that this definition needs to be extended for the case
of precessing orbits, since in that case $|C_{22}|$ will oscillate
even for spherical orbits (i.e. orbits with $r=const$). The extension
could be achieved by instead using a $|C_{22}|$ that is computed
in a coordinate system where the z-axis points along the instantaneous
orbital angular momentum.

We have also tested an eccentricity definition based on the coordinate
angular velocity $\omega$. 
Here the eccentricity is defined by~\cite{Husa:2007rh,Mroue:2010re}
\begin{equation}
\label{e_om_def}
e_{\omega}(t) = \frac{ \omega(t)-\omega_{fit}(t)}{2\omega_{fit}(t)} ,
\end{equation}
where $\omega(t)$ is simply the coordinate angular velocity
and $\omega_{fit}(t)$ is a polynomial fit of order 5 to $\omega(t)$
over a time interval corresponding to several complete orbits.
The hope is that the fit will smooth out oscillations so that
$e_{\omega}(t) \propto \omega(t)-\omega_{fit}(t)$ becomes a measure of
how much $\omega$ oscillates. The actual eccentricity is the maximum
magnitude of $e_{\omega}(t)$.

In order to compare these three eccentricity definitions we now
present the eccentricities from an actual numerical simulation.
\begin{figure}
\includegraphics[scale=0.33,clip=true]{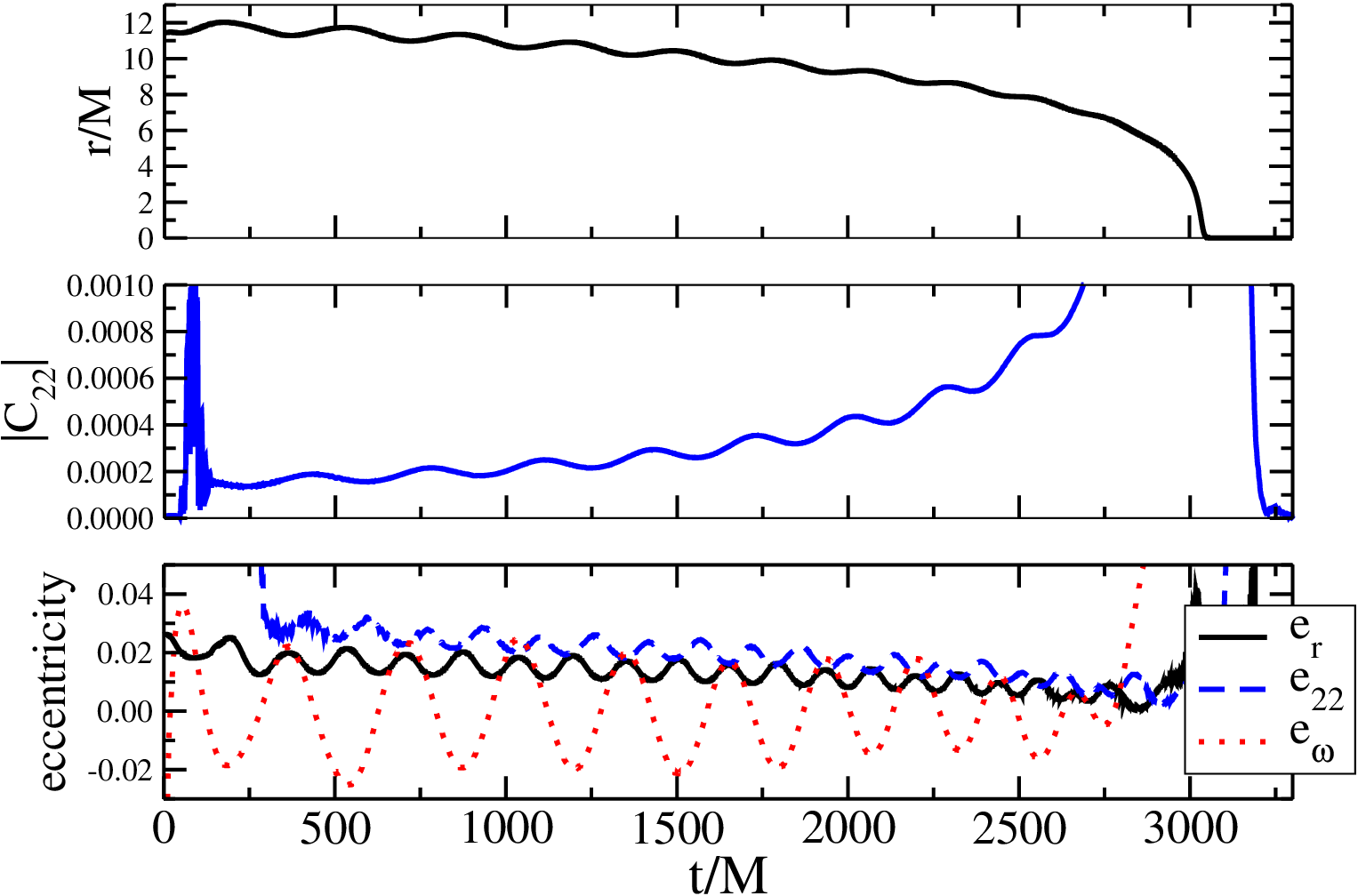}
\caption{\label{ecc_comp}
This plot shows results from a numerical run (with parameters
from row 6 of table~\ref{punc_par_tab}.
The upper panel shows the coordinate separation $r$. The middle panel
depicts the magnitude of the $l=m=2$ mode $|C_{22}|$ of $\Psi_4$
extracted at a separation of $70M$ from the center of mass.  
In the lower panel we plot the three different eccentricities obtained
with the definitions given in Eqs.~(\ref{e_r_def}), (\ref{e_22_def})
and (\ref{e_om_def}).
Note that $e_{r}$ and $e_{r}$ directly measure eccentricity,
while for $e_{\omega}$ the eccentricity corresponds to the
maximum values of $|e_{\omega}(t)|$.
}
\end{figure}
As we can see in Fig.~\ref{ecc_comp}
all three eccentricity definitions agree well
for $300M \leq t \leq3000M$. They yield a value of about $0.02$.
Notice that $e_{r}$ and $e_{22}$ are direct eccentricity measures,
while in the case of $e_{\omega}$ the eccentricity corresponds to the
maximum value of the magnitude of $e_{\omega}$.
The eccentricity definition $e_{r}$ which is calculated from the
separation has certain problems for $t<T/2\sim135M$. These occur
because we have no data for $t<0$, 
which is needed in the average over one complete period (centered around
$t$) and also for the fitting in Eq.~(\ref{r_lin_fit}).
These same problems also affect $e_{22}$. In $e_{22}$, however, they
are exacerbated by the initial junk radiation
that dominates $|C_{22}|$ until about $150M$ (see middle panel).
We see that $e_{22}$ does not completely settle down until about $700M$.
Notice also that during standard moving puncture evolutions the 
coordinates adjust quite rapidly initially. This means that any eccentricity
definition that is based on the coordinate separation or the
coordinate angular velocity will not be completely reliable during the
first $100M$ or so. Thus it comes as no surprise that the eccentricity
definition $e_{\omega}$ has a different frequency and
amplitude in the beginning.

The curves for $e_{r}$ and $e_{22}$ in Fig.~\ref{ecc_comp} oscillate
even at later times. This oscillation is due to the fact that the period $T$
which we get from Kepler's law is not exactly equal to the actual orbital 
period. The magnitude of this oscillation can be used as an error estimate
of our eccentricity measures. For a conservative eccentricity estimate
we can use the maximum values of $e_{r}$ and $e_{22}$.

Let us point out that $e_{r}$ and $e_{22}$ are easier to compute
than $e_{\omega}$. The latter depends on a polynomial fit to the measured
orbital angular velocity. The problem is that this fit has to be done over
a certain time interval, that must terminate well before the merger.
Thus it requires a certain amount of fine tuning and human intervention.
On the other hand $e_{r}$ and $e_{22}$ can be computed at any time in a
very simple way. Recall however, that $e_{22}$ is more sensitive
to the initial junk radiation. Thus for short runs used to 
probe the eccentricity of a configuration we usually
just use $e_{r}$.

\section{Numerical evolutions}
\label{numerics}

The numerical results discussed in this paper have been obtained with the
BAM code~\cite{Bruegmann:2003aw,Bruegmann:2006at,Marronetti:2007wz}.
As already mentioned, the gravitational fields are evolved using the
BSSNOK formalism~\cite{Nakamura87,Shibata95,Baumgarte:1998te} in the
variation known as the 
``moving punctures" method~\cite{Campanelli:2005dd, Baker:2005vv}.
The particulars of our BSSNOK implementation can be 
found in~\cite{Bruegmann:2006at, Marronetti:2007wz}. 
For completeness we note that lapse and shift evolve according to
\ba
(\partial_t - \beta^i\partial_i) \alpha &=& -2\alpha K , \nonumber \\
(\partial_t - \beta^k\partial_k)\beta^i &=& \frac{3}{4} B^i , \nonumber \\
(\partial_t - \beta^k\partial_k) B^i    &=&
  (\partial_t - \beta^k\partial_k)\tilde \Gamma^i - \eta B^i . 
\ea
The shift driver parameter is set $\eta =2/M$ in all our runs.

The BAM code is based on a method of lines approach using 
sixth order finite differencing in space and 
explicit fourth order Runge-Kutta time stepping.
The time step size is chosen such that the Courant factor is either
$0.25$ or $0.5$. For efficiency, Berger-Oliger type mesh refinement is
used~\cite{Berger84}. 
The numerical domain is represented by a hierarchy of nested Cartesian
boxes. The hierarchy consists of $L+1$ levels of refinement,
indexed by $l = 0, \ldots, L$. A refinement level consists
of one or two Cartesian boxes with a constant grid-spacing 
$h_l = h_0/2^l$ on level $l$. We have used here $L=10$ to $11$ for the 
number of refinement levels, with the levels 0 through 5 each 
consisting of a single fixed box centered on the
origin (the center of mass). On each of the finer levels 6 through 
$L$, we initially use two sets of moving boxes centered on each 
black hole. When the black holes get close enough that
two of these boxes start touching, they are replaced by a single
box. The position of each hole is tracked by integrating 
the shift vector. We use this same set up but with 
different resolutions depending on the purpose of each simulation.

For an accurate simulation of the inspiral and merger of
two non-spinning equal mass black holes
we might use $L=10$ with a resolution $h_{10}=M/96$ on
the finest level using 144 points on the fixed levels and 72 points on the
moving levels.
The notation we use to describe this grid setup for this simulation is:
\begin{equation}
[5\times 72, 6\times 144][M/h_{10}=96, OB=768M][C=0.25]
\end{equation}
which indicates that we have 5 moving levels with 72 points in each box and
6 fixed levels with 144 points each. The resolution is given 
by $M/h_{10}=96$ on the finest level, which results in an outer 
boundary at $768M$. 
The Courant factor here is chosen to be $0.25$, which implies
a time step of $dt_{10}=0.25 h_{10}=M/384$ on the finest level.
If the black holes have spins and/or unequal masses even more resolution
is needed. For example for a mass ratio of 3 and a dimensionless spin
magnitude of $0.6$ on the larger hole we might use a setup described by:
\begin{equation}
[6\times 72, 6\times 144][M/h_{11}=192, OB=768M][C=0.25]
\end{equation}
this setup has twice the resolution on the finest level, so that the
resolution in terms of the individual masses is now given by
$m_1/h_{11}=48$ for the smaller and $m_2/h_{11}=144$ for the larger hole.
Such runs are quite expensive. Until merger they take about one month on a
Cray XT5 supercomputer like NICS Kraken if we use 96 cores.

However, if our objective is to simply measure the orbital 
eccentricity of a binary (characterized by certain initial parameters),
it is sufficient to evolve for only a few orbits. We have found
that such an evolution does not need great accuracy. So if the goal
is to simply determine the initial eccentricity we use the following setup
\begin{equation}
\label{ecc_det_run}
[5\times 48, 6\times 96][M/h_{10}=85.3, OB=576M][C=0.5]
\end{equation}
With this grid setup it takes only two days to evolve up $t=800M$
if we use 48 cores on NICS Cray XT5 Kraken.
\begin{figure}
\includegraphics[scale=0.33,clip=true]{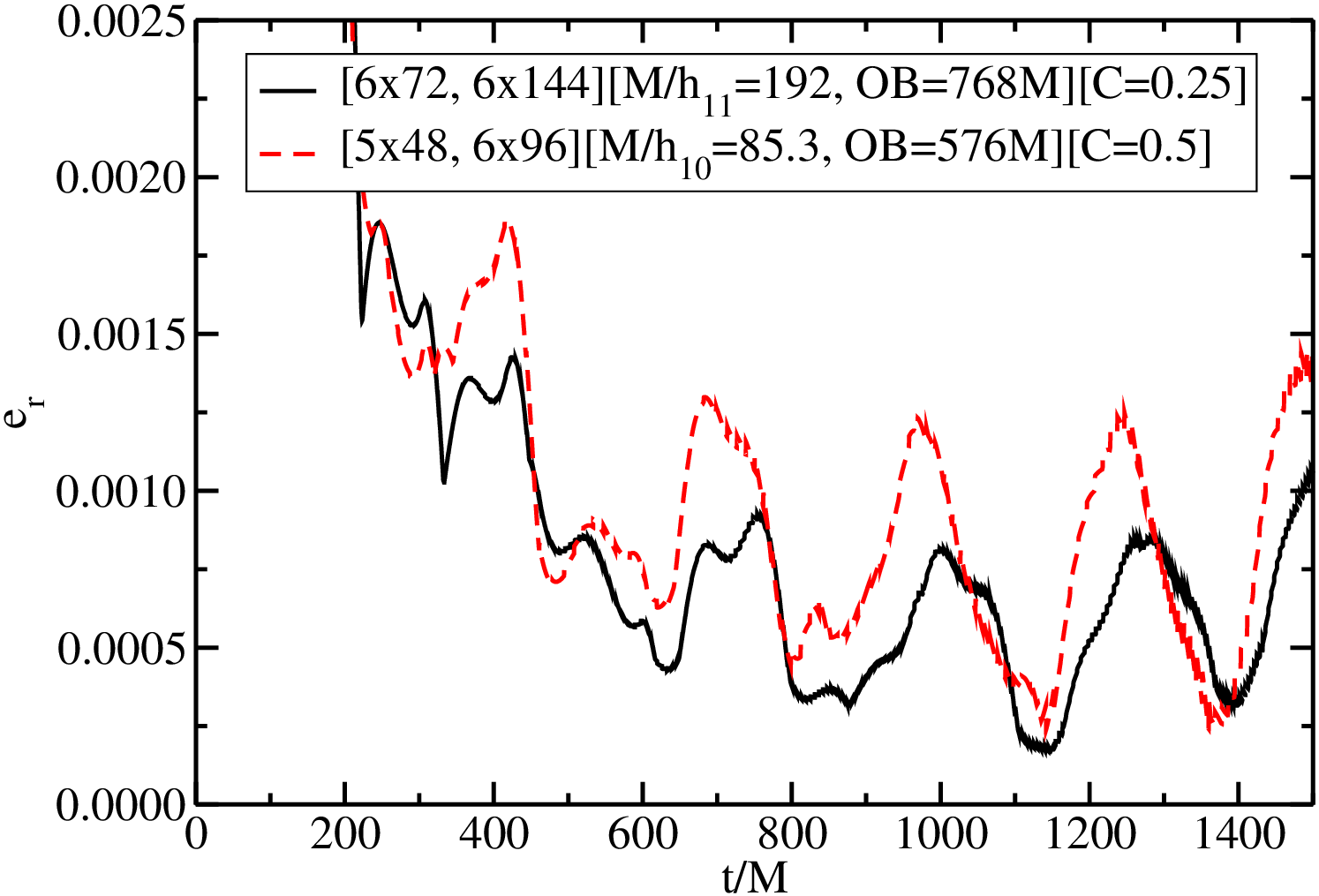}
\caption{\label{er72_VS_48h12l10cfp5}
This plot shows the eccentricity $e_r$ for two different grid setups. Both 
start with the parameters from row 7 of table~\ref{punc_par_tab}.
The solid line shows $e_r$ for a high resolution grid, while the
broken line shows the measured $e_r$ if we use a less accurate and
computer intensive grid setup. Both yield similar orbital eccentricities.
}
\end{figure}
In Fig. \ref{er72_VS_48h12l10cfp5} we see that the eccentricity $e_r$
from this cheaper run agrees quite well with the $e_r$ from a more expensive
run. As explained above no eccentricity estimate is accurate at
the start of a run. We need to wait until about $500M$ for $e_r$ to settle
down to a regular oscillation.
As mentioned before we use the maxima of $e_r$ as a conservative
eccentricity estimate. Thus the eccentricities of the cheap and expensive
runs are $0.0013$ and $0.0010$. Hence we need
to evolve until at least about $800M$ to get a reliable estimate.
From the oscillations in the curves we see that the errors in both
eccentricity estimates are about $0.0005$.

\section{Choosing puncture parameters}
\label{parchoices}

In order to start our simulations we need initial data for 
binary black holes with arbitrary spins and masses at some given
initial separation $r$. Since we will employ the moving punctures approach
in our evolutions, we will use standard puncture 
initial data~\cite{Brandt97b}. Thus the 3-metric and extrinsic curvature
are given by
\begin{eqnarray}
\label{punc-3-metric}
g_{ij} &=& \psi^4 \delta_{ij} \\
K^{ij} &=& \psi^{-10} \sum_{A=1}^2 \Big\{
   \frac{3}{r_A^2}\left[
      p_A^{(i} n_A^{j)} 
    - \frac{\delta_{kl}p_A^k n_A^l}{2}(\delta^{ij}-n_A^i n_A^j)
   \right] \nonumber \\
&&+\frac{6}{r_A^2}\left[\epsilon_{klm} S_A^l n_A^m \delta^{k(i} n_A^{j)}
   \right]
\Big\} .
\end{eqnarray}
Here $p_A^i$ and $S_A^i$ are the momentum and spin parameters of
black hole $A$, while $r_A$ and $n_A^i$ denote the distance and normal
vector measured from hole $A$.
The conformal factor is
\begin{equation}
\label{psi_punc}
\psi = 1 + \frac{m_{b_1}}{2r_1} + \frac{m_{b_2}}{2r_2} + u ,
\end{equation}
where $m_{b_1}$ and $m_{b_2}$ denote the black hole bare mass parameters.
The scalar $u$ is computed by numerically solving the Hamiltonian
constraint.
These initial data are very flexible since the free parameters
for the position, momentum and spin of each black hole can be chosen
freely. Thus one can setup practically any kind of orbit.
Note, however, that our goal is
to set up data for black holes that are in quasi-circular orbit. This
means that we need to choose our momentum parameters such that the
eccentricity is as small as possible. Since we start our evolutions
in a frame where the center of mass is at rest, both black holes
have momenta that are equal in magnitude but opposite in direction.
Thus we have to choose only two parameters: the tangential and radial
component of the momentum of one of the black holes.

To complete the definition of the initial data, we also need to specify
initial values for the lapse $\alpha$ and shift
vector $\beta^i$. At time $t=0$ we use
\begin{eqnarray}
\label{ini_lapse_shift}
\alpha &=& \psi^{-2}, \nonumber\\
\beta^i &=& 0 .
\end{eqnarray}

\subsection{Finding the momentum parameters}
\label{finding_p_pars}

There have been various attempts to guess appropriate
momentum parameters. Some have used the quasi-equilibrium 
approach~\cite{Baumgarte00a,Tichy:2003zg,Tichy:2003qi}, but most are based
on post-Newtonian (PN) 
approximations (see
e.g.~\cite{Marronetti:2007wz,Marronetti:2007ya,Brugmann:2008zz}).
The latter have been taken to their extreme
in~\cite{Husa:2007rh} and~\cite{Walther:2009ng} who integrate PN
equations in the ADMTT gauge~\cite{Schaefer85} in the hope 
that the thus obtained momentum parameters will lead to 
orbits with less eccentricity. 
However, as mentioned in~\cite{Walther:2009ng},
it is not certain that non-eccentric PN parameters in ADMTT gauge 
should produce non-eccentric orbits in full General Relativity.
We want to point out here,
that standard puncture data are inconsistent with PN theory beyond
$(v/c)^3$
~\footnote{The standard PN jargon for a term of order $(v/c)^n$ in
the equations of motion is $\frac{n}{2}$PN term. Notice that according
to the virial theorem $v^2\sim M/r$, so that $\frac{n}{2}$PN terms
are also $O(M/r)^{\frac{n}{2}}$.}
~\cite{Tichy02,Yunes:2006iw,Yunes:2005nn,Marronetti:2007ya},
even in ADMTT gauge!
The reasons for this inconsistency are as follows. First,
the 3-metric of puncture data is always conformally
flat (see Eq.(\ref{punc-3-metric})), 
while the PN 3-metric contains deviations
from conformal flatness at order
$(v/c)^4$~\cite{Tichy02,JohnsonMcDaniel:2009dq}.
This is true for both harmonic and ADMTT gauge.
Second, the conformal factor in ADMTT gauge is given by~\cite{Tichy02}
\begin{equation}
\label{psi_PN}
\psi_{PN} = 1 + \frac{E_1}{2r_1} + \frac{E_2}{2r_2} ,
\end{equation}
where $E_A$ is the energy of each particle used to model the black holes.
If we compare Eqs.~(\ref{psi_PN}) and (\ref{psi_punc}) we see that
the conformal factor in ADMTT gauge is not identical to the
conformal factor used in puncture initial data.
One difference is that the ADMTT $\psi_{PN}$ contains the particle
energies $E_A$ while the puncture $\psi$ contains the bare masses $m_{b_A}$.
These two only agree for infinite separation. Furthermore, the puncture
$\psi$ contains the additional piece $u$ that is obtained by
numerically solving the Hamiltonian constraint, while PN data
violates the Hamiltonian constraint. One might like to argue that
the PN data should agree with puncture data up to the higher order terms
that are neglected in the PN approximation. This is however not true
close to the black holes, because PN theory is not valid in regions of
very strong gravity. Thus near the black holes the ADMTT metric differs
from the puncture metric by terms that are of low PN order.
 
From the above explanation it is clear that the coordinate
separation in ADMTT gauge does not have the same physical meaning
as the coordinate separation in puncture initial data.
Thus if by some method we arrive at the momentum
parameters needed at a particular separation $r$ (in ADMTT coordinates),
we should not simply use these same momentum parameters at the same
puncture separation.

So if (for some configurations) the more complicated
approaches in~\cite{Husa:2007rh} or~\cite{Walther:2009ng}
lead to less eccentric orbits than e.g. the simpler approaches
in~\cite{Marronetti:2007wz,Marronetti:2007ya,Brugmann:2008zz}
this should be considered a coincidence, since it is
not due to the inclusion of higher order terms or due to
the integration of post-Newtonian equations of motion. 
In fact, we find that (as anticipated by~\cite{Walther:2009ng})
for many configurations
the approaches in~\cite{Husa:2007rh} or~\cite{Walther:2009ng}
do not lead to less eccentricity than the ones 
in~\cite{Marronetti:2007wz,Marronetti:2007ya,Brugmann:2008zz}.

For these reasons we take a different approach here. We will take
a simple PN formula to obtain a reasonable guess for both
the tangential and radial momentum components $p_t$ and $p_r$. 
Then we numerically evolve the resulting initial data for a short time
(using the efficient grid setup in Eq.~(\ref{ecc_det_run}))
to see how eccentric they really are. Afterward we adjust $p_t$ to reduce
the eccentricity. In this way we can obtain low eccentricity orbits
for any configuration.
In order to come up with a reasonable guess for $p_t$ and $p_r$
we use 2PN accurate expressions of Kidder~\cite{Kidder:1995zr}
in harmonic gauge.
Specifically, we freely choose the two masses $m_1$, $m_2$, the
six spin components of $\mathbf{S}_1$ and $\mathbf{S}_2$ 
and a separation $r$.
Next, choosing $\hat{\mathbf{L}}_N$ in the $z$-direction,
we use Eqs.~(2.8) and (4.7) of~\cite{Kidder:1995zr} to compute
the total orbital angular momentum $\mathbf{L}$, and 
Eq.~(4.12) of~\cite{Kidder:1995zr} to compute
$\dot{r}$. We rotate $\mathbf{L}$, $\mathbf{S}_1$ and $\mathbf{S}_2$
so that $\mathbf{L}$ points in the $z$-direction. Then we obtain
the momentum in the $xy$-plane as
\begin{eqnarray}
\label{PN_mom_pt}
p_t = |\mathbf{L}|/r  \\
\label{PN_mom_pr}
p_r = \mu |\dot{r}| .
\end{eqnarray}
In all cases discussed in this paper we put the two punctures 
on the $y$-axis at
$y_1 =  m_2 r/M$ and 
$y_2 = -m_1 r/M$.
The two initial black hole momenta are then 
$\mathbf{p}_{1}=(-p_t,-p_r,0)$ and
$\mathbf{p}_{2}= (p_t,p_r,0)$.

\subsection{Determining the bare mass parameters}

The momenta $\mathbf{p}_{1}$, $\mathbf{p}_{2}$ 
and the spins $\mathbf{S}_1$, $\mathbf{S}_2$ for a coordinate separation $r$  
directly enter the Bowen-York extrinsic curvature of standard puncture data.
Note, however, that the bare mass parameters $m_{b_1}$ and $m_{b_2}$
which appear in the construction of standard puncture data are not equal to
the individual black hole masses. As
in~\cite{Tichy:2007hk,Marronetti:2007wz,Marronetti:2007ya,Tichy:2008du}
we obtain the bare masses from the condition that
the ADM masses 
\begin{equation}
m^{ADM}_{A} = m_{b_A} (1+ u_A) +  \frac{m_{b_1} m_{b_2}}{2r}  ,
\end{equation}
measured at the punctures~\cite{Tichy03a} 
should be equal to $m_1$ and $m_2$, where $u_A$ is the value of $u$
at puncture $A$.
As in~\cite{Tichy03a,Tichy:2003qi,Ansorg:2004ds} we assume that the ADM
masses measured at each puncture are a good approximation for the
initial individual black hole masses. Numerically this condition
is implemented as a root finder in the initial data solver that
picks $m_{b_1}$ and $m_{b_2}$ such that the ADM masses at the punctures 
are equal to $m_1$ and $m_2$.

\section{Reducing the eccentricity}
\label{ecc_reduc}

Now that we know how to generate initial data on approximately
circular orbits for arbitrary spins, masses and separations, it is
time to present some numerical results to demonstrate what eccentricities we
get and how we can reduce them.
For our purposes here, we consider the eccentricity small enough if it is
$0.003$ or less, which agrees with the NRAR target 
of an eccentricity of a few times $10^{-3}$.
\begin{table*}
{\small
\begin{tabular}{r|c|c|c|c|c|l|l|c|c|c}
row 
& $\frac{r}{M}$ 
& $\frac{m_{b_1}}{M}$ & $\frac{m_{b_2}}{M}$
& $\frac{m_1}{M}$ & $\frac{m_2}{M}$ 
& $\frac{|\mathbf{S}_1|}{m_1^2},\theta_1,\phi_1$
& $\frac{|\mathbf{S}_2|}{m_2^2},\theta_2,\phi_2$
& $\frac{10^2 p_t}{M}$   & $\frac{10^4 p_r}{M}$ & $10^3 e_r$ \\
\hline
%
1 & 11.9718 & 0.488255 & 0.488255 & 1/2 & 1/2  & $0    $    & $0$        & 8.5018 & 3.88 & 5.0 \\
2 & 11.9718 & 0.488249 & 0.488249 & 1/2 & 1/2  & $0$        & $0$        & 8.5168 & 3.88 & 1.5 \\
\hline
3 & 11.9694 & 0.404597 & 0.404687 & 1/2 & 1/2  &$0.6,60,0$  &$0.6,120,90$& 8.5013& 3.87& 4.0 \\
4 & 11.9694 & 0.404592 & 0.404681 & 1/2 & 1/2  &$0.6,60,0$  &$0.6,120,90$& 8.5163& 3.87& 2.0 \\
5 & 11.9694 & 0.404588 & 0.404677 & 1/2 & 1/2  &$0.6,60,0$  &$0.6,120,90$& 8.5280& 3.87& 1.5 \\
\hline
6 & 11.4546 & 0.2239\phantom{00} & 0.6145\phantom{00} & 1/4 & 3/4 & $0.4,0,0$  & $0.6,0,0$  & 6.4416 & 2.21 & 20 \\
7 & 10.7271 & 0.223341 & 0.614005 & 1/4 & 3/4 & $0.4,0,0$  & $0.6,0,0$  & 6.6396 & 2.50 & 1.0 \\
\hline
%
8 & 12.0815 & 0.224174 & 0.614794 & 1/4 & 3/4 & $0.4,180,0$ & $0.6,180,0$ & 6.5900 & 2.50 & 1.0 \\
9 & 11.7173 & 0.223891 & 0.614572 & 1/4 & 3/4 & $0.4,180,0$ & $0.6,180,0$ & 6.7497 & 4.08 & 5.0 \\
\hline
%
10 & 10.9876 & 0.223317 & 0.614121 & 1/4 & 3/4 & $0.4,0,0$  & $0.6,180,0$ & 6.9756 & 3.19 & 3.0 \\
11 & 11.5010 & 0.300017 & 0.543886 & 1/3  & 2/3  & $0.4,60,0$ & $0.6,60,0$  & 7.6504 & 3.18 & 2.0 \\
12 & 11.5033 & 0.300020 & 0.543926 & 1/3  & 2/3  & $0.4,60,0$ & $0.6,60,90$ & 7.6509 & 3.19 & 1.8 \\
13 & 11.6182 & 0.300023 & 0.543980 & 1/3  & 2/3  & $0.4,60,0$ & $0.6,120,90$& 7.7896 & 3.51 & 2.1 \\
14 & 11.5502 & 0.310095 & 0.656229 & 1/3  & 2/3  & $0.3,0,0$  & $0$         & 7.7196 & 3.32 & 1.3 \\
\end{tabular}
}
\caption{\label{punc_par_tab}
Initial data parameters. The black holes have coordinate
separation $r$. We give both bare masses $m_{b_1}$, $m_{b_2}$ as well
as physical masses $m_1$, $m_2$.
The punctures are located on the
$y$-axis at $y_1 = m_2 r/M$ and $y_2 = -m_1 r/M$.
The spins are given in terms of their magnitudes and the usual polar
angles of spherical coordinates measured in degrees. 
The linear momenta are $(\mp p_t, \mp p_r, 0)$. The last column
shows the resulting eccentricity.
}
\end{table*}
The first example we consider is an equal mass binary without spin.
The momentum parameters are picked according to Eqs.~(\ref{PN_mom_pt}) 
and (\ref{PN_mom_pr}). The values of all the initial
parameters are given in the first line of table \ref{punc_par_tab}.
As we can see, the eccentricity is about 0.005 in this case. In order
to test how it varies with $p_t$ we have also performed a run where
we have increased $p_t$ by
\begin{equation}
\label{p_corr}
\Delta p_t = \mu \sqrt{\frac{M}{r}}
\left[  11.29\left(\frac{M}{r}\right)^3 
      - 92.37\left(\frac{M}{r}\right)^4\right] .
\end{equation}
The results of this increase are shown in the second line of table
\ref{punc_par_tab} and yield an eccentricity that is reduced by more than a
factor of 3. The expression for Eq.~(\ref{p_corr}) comes from the fitting
of two of our older equal mass runs that resulted in low eccentricity.
Thus non-zero spins or unequal masses are not taken into account
by this fit. Therefore Eq.~(\ref{p_corr}) certainly does
not present the optimum momentum correction.
In fact we have found that adding $\Delta p_t$ to $p_t$ does not
always reduce the eccentricity since the PN
estimate of Eq.~(\ref{PN_mom_pt}) for $p_t$ is sometimes too
large and sometimes too small for generic orbits. 
The expression in Eq.~(\ref{p_corr}) simply
gives us a rough estimate by how much we might have
to raise or lower $p_t$ in order to reduce the eccentricity. 

So in order to really reduce the eccentricity we usually start
one run on the coarse grid described by Eq.~(\ref{ecc_det_run}).
This run normally uses 
$p_t$ given by Eq.~(\ref{PN_mom_pt}). We then look at the coordinate
separation $r(t)$ for this run. Usually one can tell whether the
initial tangential momentum $p_t$ was too large or too small. We then
start a new run where we either increase or decrease 
$p_t$ by $\Delta p_t$. This run then gives a new $r(t)$ and $e_r$.
From the two resulting simulations one can then extrapolate
to zero eccentricity to obtain a more refined tangential momentum
parameter. This procedure is illustrated in rows 3,4, and 5 of
table~\ref{punc_par_tab}. In row 3 we have used the 
$p_t=8.5013\times 10^{-2} M$ from Eq.~(\ref{PN_mom_pt}) and
obtained an eccentricity of 0.004.
In row 4 we have increased this $p_t$ by $\Delta p_t$ which leads
to a decrease of the eccentricity to 0.002. A further increase to
$p_t=8.5280\times 10^{-2} M$ (as in row 5) then yields an
eccentricity of 0.0015.

Row 6 of table~\ref{punc_par_tab} shows the initial data that were
used to produce Fig.~\ref{ecc_comp}. In order to produce an eccentricity
which is clearly visible in the $r(t)$ curve we have
used a $p_t$ that is deliberately chosen much larger than
what is predicted by Eq.~(\ref{PN_mom_pt}). In row 7 we show the
same configuration (at somewhat closer separation), but this
time we choose $p_t$ according to Eq.~(\ref{PN_mom_pt}), we see that in
this case the eccentricity is already small enough to satisfy e.g.
the NRAR guidelines.

In row 8 we have used our method to produce momentum parameters
for a similar configuration, but this time both spins point in the negative
$z$-direction. Again we can reach the NRAR target of an eccentricity
of order 0.001. In row 9 we also
show the eccentricity resulting from the method introduced
in~\cite{Walther:2009ng} for the same mass ratio and spin configuration.
We can see that it is about 5 times larger, which supports our argument
that integrating PN equations of motion does not necessarily lead
to better results. 
\begin{figure}
\includegraphics[scale=0.33,clip=true]{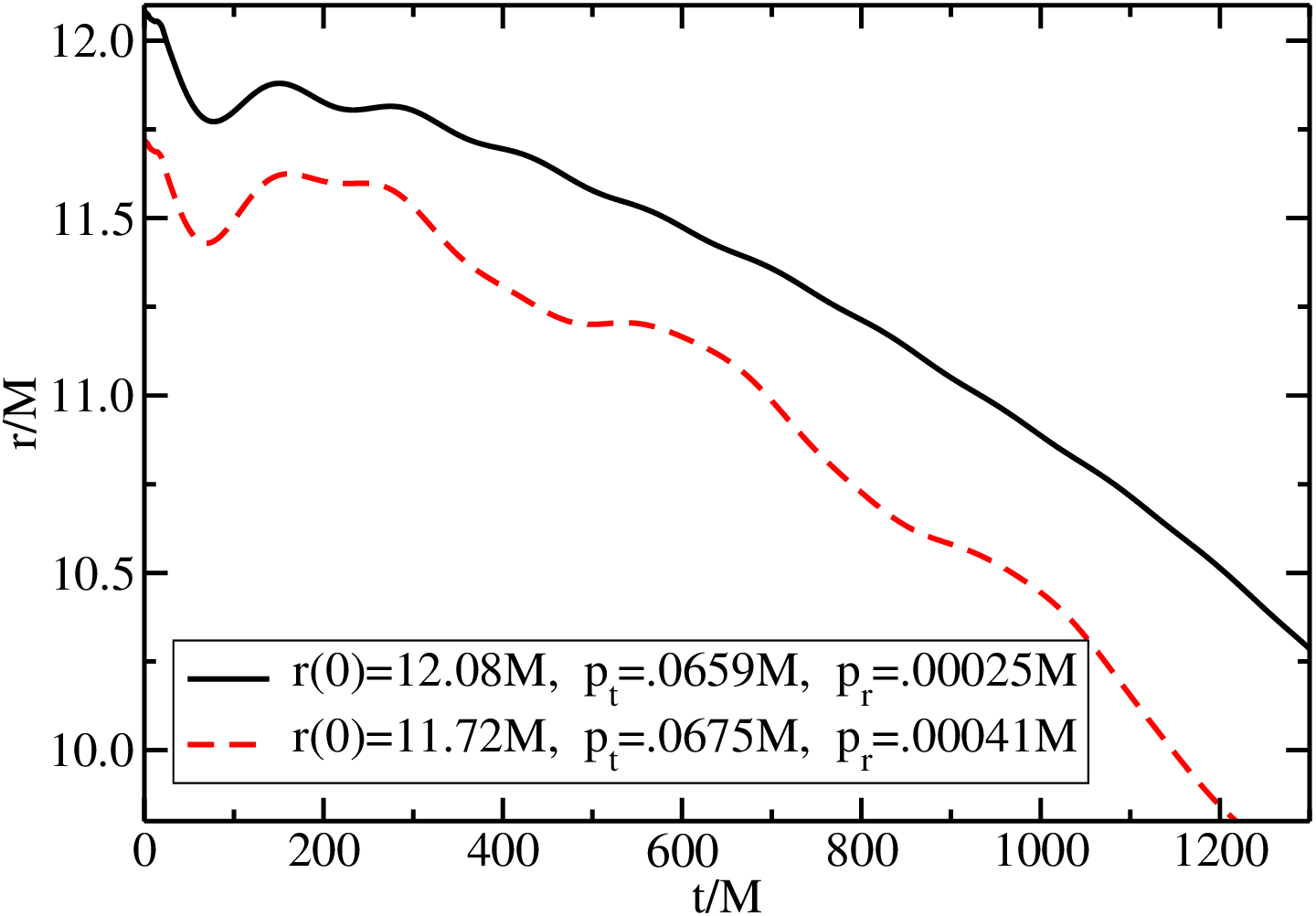}
\caption{\label{r_WTvsPM}
This plot shows $r(t)$ for the two numerical runs with parameters
from rows 8 (solid line) and 9 (broken line) of table~\ref{punc_par_tab}.
The parameters for the solid line were picked according to the method
explained in this paper while the broken line is the result of
setting the parameters according to the
much more complicated method in~\cite{Walther:2009ng}.
As we can see the eccentricity difference of about a factor of 5
is quite noticeable in the $r(t)$ curves.
}
\end{figure}
The results from these two runs are also shown in Fig.~\ref{r_WTvsPM}.
The solid line corresponds to row 8 (i.e. our approach) and the broken
line is generated using the approach in~\cite{Walther:2009ng}. As we can
see our method does give a noticeably less eccentric $r(t)$ curve.
Notice that the initial dip in the coordinate separation $r(t)$,
is due to the aforementioned initial coordinate adjustment.
It does not mean the two holes really plunge toward each other initially.

The last five rows of table~\ref{punc_par_tab} give a few more
examples of parameters and resulting eccentricities. They show
that we can obtain low eccentricities for generic mass ratios and spin
orientations.

Notice that our eccentricity reduction method is similar in spirit to
the method described in~\cite{Pfeiffer:2007yz,Boyle:2007ft}, in that we
also use short numerical runs to adjust certain parameters. There are,
however, important differences. In~\cite{Pfeiffer:2007yz,Boyle:2007ft}
excision type initial data, not punctures are used. These data
are constructed using an extension to the conformal thin sandwich 
formalism~\cite{York99}.
Empirically it turns out that the tangential momentum that is
achieved with the conformal thin sandwich method is quite close to what
is needed for low eccentricity data. The main reason for eccentricity
is the absence of a radial momentum in the original conformal thin sandwich
method. In~\cite{Pfeiffer:2007yz} a method is developed that adds
an arbitrary radial velocity parameter to the initial data.
This radial velocity parameter together with the tangential momentum
are then adjusted to reduce the eccentricity. The method
introduced in~\cite{Pfeiffer:2007yz,Boyle:2007ft} is capable of 
producing eccentricities of order of only $10^{-5}$.
For standard puncture data the method
in~\cite{Pfeiffer:2007yz,Boyle:2007ft} cannot be used directly. 
One problem is that standard moving puncture simulations start with
the lapse and shift given in Eqs.~(\ref{ini_lapse_shift}). Thus the
coordinates used are not well adapted to quasi-equilibrium. Hence
oscillations in the hole separation $r$ are not due to real eccentricity
alone but also due to the fact that the coordinates are still evolving
as well. This problem is quite visible in Fig.~\ref{r_WTvsPM}. The broken
line shows oscillations that have more than one frequency.
Thus we cannot fit the curve very well with a straight line plus a single 
sine function as in~\cite{Pfeiffer:2007yz,Boyle:2007ft}.
To summarize, reducing the eccentricity for puncture data is harder than for
the extended conformal thin sandwich data in~\cite{Pfeiffer:2007yz}.
Therefore, our results in both final eccentricities and 
reduction of eccentricity per iteration are not as good as
in~\cite{Pfeiffer:2007yz,Boyle:2007ft}.

In the cases we have studied so far, we have found it unnecessary to
adjust $p_r$ (away from the PN value in Eq.~(\ref{PN_mom_pr})) 
in order to reach an eccentricity of order 0.001. 
It is much more important to choose an appropriate $p_t$.
However, we expect that adjusting $p_r$ will be necessary 
to reach even lower eccentricities.

\section{Discussion}
\label{discussion}

We have introduced the two new eccentricity measures $e_r$ and $e_{22}$.
Both are easy to compute since their calculation does not involve any
free parameters unlike earlier definitions akin to $e_{\omega}(t)$.
To compute eccentricity measures such as $e_{\omega}(t)$ one
needs to specify a time interval during the inspiral phase over which
we fit the orbital angular velocity to a polynomial of some low degree.
This degree is essentially another free parameter and is usually
chosen to be 4 or 5.
Note, however, that all eccentricity definitions are ambiguous to a certain
extend since the entire concept of eccentricity is only rigorously defined
for periodic orbits. All eccentricity definitions for inspiral orbits depend
on how we split a function of time (like the separation) into a smooth and
an oscillating piece. Thus we do not claim that our definitions 
$e_r$ and $e_{22}$ are less ambiguous than earlier ones. 
For example our definitions depend on the period $T$ which we compute
from Kepler's law, but other choices are possible (e.g. an estimate of
the actual orbital period of the numerical simulation at each time).
Since our definitions $e_r$ and $e_{22}$ do not require us to choose 
extra parameters such as fitting intervals they are easier to use.

We also show that certain low resolution grid setups (which are
relatively inexpensive) can be used to estimate the initial eccentricity
using our measure $e_r$. This gives us a relatively efficient
way to measure the eccentricity of any initial data.

Furthermore, we have explained why the coordinates of 
standard puncture data are not the same as in PN calculations
in ADMTT or harmonic gauge. Thus even if one arrives at a highly
accurate estimate of the momentum parameters needed for low eccentricity
orbits in ADMTT gauge, there is no easy way to incorporate this knowledge
into simulations starting from standard puncture data. 
Indeed we find that using accurate parameters in ADMTT
gauge~\cite{Husa:2007rh,Walther:2009ng} in standard puncture initial data
can lead to relatively large eccentricities.

We provide a simpler approach which starts from momentum parameters 
using the relatively short expressions given by Kidder~\cite{Kidder:1995zr}.
After measuring the resulting eccentricity $e_r$ in a short inexpensive
numerical simulation, these parameters can then be refined by
changing $p_t$ by an amount of order $\Delta p_t$ (see Eq.~(\ref{p_corr})).
In this way we can always arrive at eccentricities that are low enough for
the purposes of the NRAR collaboration~\cite{NRAR_web}.
In order to achieve this we usually do not need more than three
numerical runs.

\begin{acknowledgments}

It is a pleasure to thank Doreen M\"uller for useful discussions about the 
approach in~\cite{Walther:2009ng}.
This work was supported by NSF grant PHY-0855315.
Computational resources were provided by the Ranger cluster at
the Texas Advanced Computing Center (allocation TG-PHY090095) 
and the Kraken cluster (allocation TG-PHY100051)
at the National Institute for Computational Sciences.

\end{acknowledgments}




\bibliography{references}

\end{document}